\documentclass[twocolumn,showpacs,preprintnumbers,amsmath,amssymb]{revtex4}

\usepackage{graphicx}
\usepackage[latin1]{inputenc}
\usepackage{dcolumn}
\usepackage{bm}
\usepackage{epsfig}
\usepackage{color}
\usepackage{amsthm}

\newtheorem{theorem}{Theorem}

\newtheorem{proposition}{Proposition}

\begin{document}


\title{Canonical Horizontal Visibility Graphs are uniquely determined by their degree sequence}

\author{Bartolo Luque$^1$, Lucas Lacasa$^2$}
\email{l.lacasa@qmul.ac.uk}
\affiliation{$^1$Department of Applied Mathematics, School of Aeronautics, Technical University of Madrid (UPM), Plaza Cardenal Cisneros 28040 Madrid (Spain)\\$^2$School of Mathematical Sciences, Queen Mary University of London, Mile End Road London E14NS (UK)}%

\begin{abstract}
Horizontal visibility graphs (HVGs) are graphs constructed in correspondence with number sequences that have been introduced and explored recently in the context of graph-theoretical time series analysis.  In most of the cases simple measures based on the degree sequence (or functionals of these such as entropies over degree and joint degree distributions) appear to be highly informative features for automatic classification and provide nontrivial information on the associated dynamical process, working even better than more sophisticated topological metrics. It is thus an open question why these seemingly simple measures capture so much information. Here we prove that, under suitable conditions, there exist a bijection between the adjacency matrix of an HVG and its degree sequence, and we give an explicit construction of such bijection. As a consequence, under these conditions HVGs are unigraphs and the degree sequence fully encapsulates all the information of these graphs, thereby giving a plausible reason for its apparently unreasonable effectiveness.
\end{abstract}


 \maketitle


The theory of horizontal visibility graphs (HVGs) \cite{pnas, pre, nonlinearity} builds a bridge between nonlinear dynamics, time series analysis and graph theory by providing a recipe to map a time series of $N$ data into a graph of $N$ vertices and subsequently studying the topological properties of the resulting graph in direct correspondence with the structure and dynamical properties of the associated series.
From a combinatoric point of view, HVGs are  outerplanar graphs with a Hamiltonian path \cite{severini}, i.e. noncrossing graphs as defined in algebraic combinatorics \cite{flajo}.
In recent years, this mapping has been successfully explored to provide a topological characterization of different routes to low dimensional chaos \cite{jns, quasi, pre2013}, or different types of stochastic and chaotic dynamics \cite{nonlinearity}. From an applied angle, this technique is being widely used to extract in a simple and computationally efficient way informative features for the description and classification of empirical time series appearing in several areas of physics including optics \cite{physics3}, fluid dynamics \cite{fluiddyn0, fluiddyn1, fluiddyn2}, geophysics \cite{physics2} or astrophysics \cite{suyal, Zou}, and extend beyond physics in areas such as physiology \cite{physio1, meditation_VG}, neuroscience \cite{neuro} or finance \cite{ryan1} to cite only a few examples. Among the wealth of possible graph-theoretical measures that one could compute on a graph, it is noticeable that the most informative metrics include the degree and joint degree distributions (as well as some moments \cite{pre2013}), entropic quantities based on these distributions or sequential motifs \cite{motifs}, all these based on the same quantity: the degree sequence. Of course the degree sequence is just one out of many possible descriptors of a graph's topology \cite{bollobas, newman}, and in this sense it is an open and relevant question why, in the context of HVG theory, this simple and computationally efficient quantity seems to be {\it enough} to describe the full complexity of a given time series.
In this letter we provide a partial solution for this question, by proving that under suitable conditions (for a general family of HVGs labelled canonical) these graphs are uniquely determined by its degree sequence, i.e.  they are so-called unigraphs \cite{uni1, uni2, uni3}. In other words, we prove that for canonical HVGs the degree sequence is the measure that encapsulates all the information of the graph. To do that, we will need to introduce a list of definitions and propositions before stating the main theorem. We will give a constructive proof of the main theorem by giving an explicit bijection between the adjacency matrix and the degree sequence, and we finally will discuss some implications of this theorem.\\

\begin{figure}
\centering
\includegraphics[width=0.8\columnwidth]{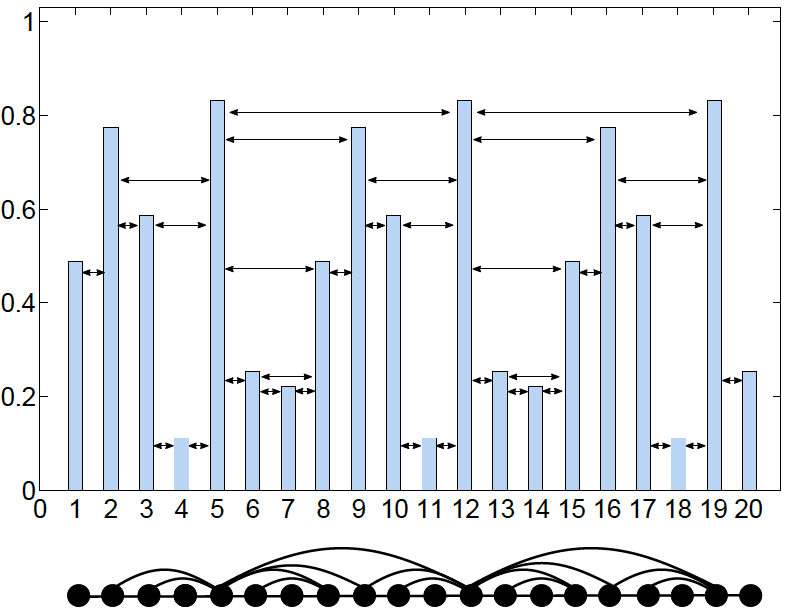}
\caption{Sample time series of 20 data and its associated horizontal visibility graph (HVG).}
\label{fig:HVG}
\end{figure}
\noindent Let ${\cal S}=\{x_1,\dots,x_N\}$, $x_i \in \mathbb{R}$ be a real-valued scalar time (or otherwise ordered) series of $N$ data. Its horizontal visibility graph HVG(${\cal S}$) is defined as an undirected graph of $N$ vertices, where each vertex $i\in [1,N]$ is labelled in correspondence with the ordered datum $x_i$. Hence $x_1$ is related to vertex $i=1$, $x_2$ to vertex $i=2$, and so on.
Then, two vertices $i$, $j$ (assume $i<j$ without loss of generality) share an edge if and only if $x_k<\inf(x_i,x_j),\ \forall k: i<k<j$. This is an ordering criterion which can be visualized in figure \ref{fig:HVG}.\\
Now, for the latter series  ${\cal S}$, if
(i) $x_1$ and $x_N$ are the two largest data in the series and (ii) the inner data are different $x_i\neq x_j \ \forall i,j=2,\dots,N-1$ then we call the series ${\cal S}$ {\it canonical} and its associated HVG is called a canonical HVG. The set of canonical HVGs is therefore a subset of all possible HVGs. The second condition is usually guaranteed for real-valued irregular time series (which are the objects under study in most of the applications), while even if the first condition is not easy to meet, for a given dynamical process it is not difficult to find a sample series of size $N$ that is generated from the dynamical process and complies with the first condition. In general this is achieved by a procedure of series canonization:\\

\begin{figure}
\centering
\includegraphics[width=1.1\columnwidth]{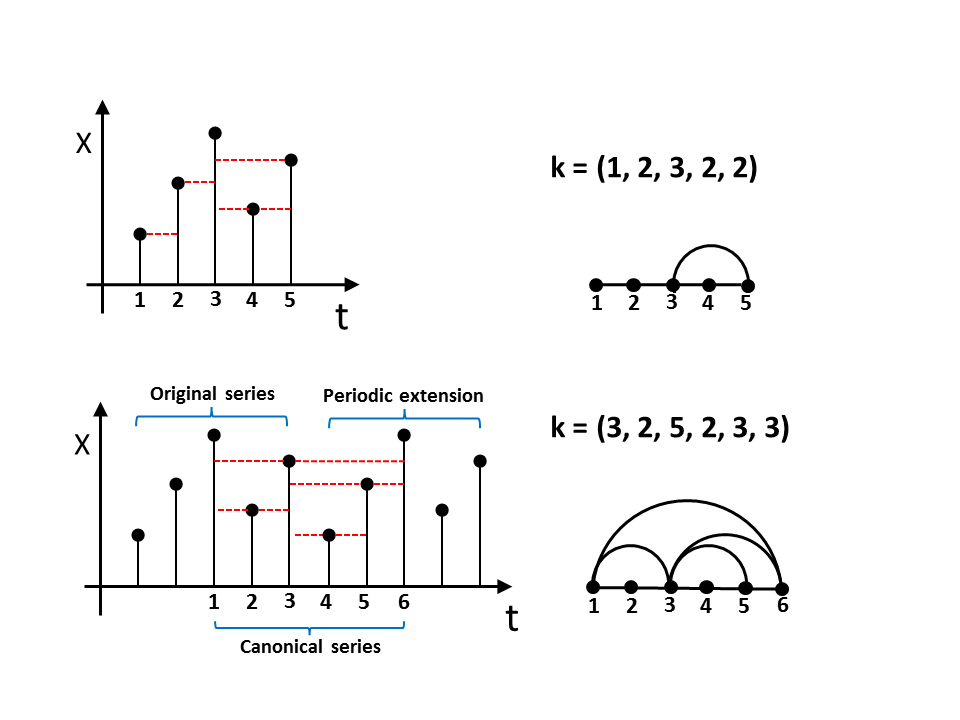}
\caption{An example of series canonization procedure.}
\label{fig:canoni}
\end{figure}

\noindent {\bf Series canonization}. Consider again a sequence ${\cal S}=\{x_1,x_2,\dots,x_N\}$ which is not canonical. One can canonize the sequence -i.e., construct an associated sequence which is canonical- using two different procedures, depending on the property that ${\cal S}$ is lacking.\\
(i) If $x_1$ and $x_N$ are not the largest data (see figure \ref{fig:canoni} for a visual example): In this case, proceed to find the integer $m$ such that $x_m$ takes the maximal value in ${\cal S}$, i.e. ${\cal S}=\{x_1,\dots,x_m,\dots,x_N\}$. Then construct the periodic extension as ${\cal S}^*=\{x_1,\dots,x_m,\dots,x_N,x_1,\dots,x_m,\dots,x_N\}$. Finally, extract from this latter sequence the subsequence of $N+1$ data ${\cal S}^c=\{x_m,\dots,x_m\}$: this is defined as the canonical series associated to ${\cal S}$.\\
(ii) If $x_{i+\tau}=x_i$ for some $\tau>0$ (the sequence is periodic of period $\tau$) but the data do not repeat within a period: Locate the largest value of the series and label it $x_m$.
Then ${\cal S}^c$ is a sequence with $\tau+1$ data formed by ${\cal S}^*=\{x_m,x_{m+1},\dots,x_{m+\tau}=x_m\}$.\\
(iii) if $x_i=x_j$ for some $i,j$ but the sequence is not periodic, or if the sequence is periodic and some data take identical values within a period, then the sequence cannot be canonized.\\
Note that any time series extracted from the trajectory of a stochastic dynamical system where the random variable takes values on $\mathbb{R}$ can be canonized using the procedure
(i) almost surely. The same holds for the trajectories of aperiodic deterministic dynamical systems. For periodic sequences, one can apply the procedure (2). In general, if a series ${\cal S}$ is extracted from a given dynamical process, then its canonized version ${\cal S}^c$ is also a possible realization of the same dynamical process with equal probability (that is, ${\cal S}$ and ${\cal S}^c$ are equiprobable microstates).\\
Before stating the main theorem, we just need to state the following:
\begin{proposition}
The HVG associated to any monotonic sequence of $N$ data is a path graph of $N$ vertices.
\label{prop}
\end{proposition}
\noindent The detailed proof of this proposition is trivial and is therefore omitted (as a sketch, one proceeds by applying the visibility criterion to the three cases -monotonically increasing, monotonically decreasing, and constant series-).
We are thus ready to state the main theorem of this paper.\\

\begin{theorem}
If $\mathcal{G}$ is a canonical HVG of $N$ vertices with adjacency matrix $\mathcal{A}$ and degree sequence ${\bf k}=(k_1,k_2,\dots,k_N)$, then there exists a bijection between $\mathcal{A}$ and $\bf{k}$. \\
\end{theorem}

\noindent {\bf Proof.} We construct this bijection explicitly. Consider the computable function $f:\{0,1\}^{N\times N}\to {\Omega}$ where $\Omega \subset [\mathbb{N}-\{0,1\}]^N$ is the subset of degree sequences that are admissible for HVGs. This function admits an inverse $f^{-1}$ that puts the adjacency matrix $\cal A$ and the degree sequence $\bf k$ of a canonical HVG in bijection. Since $f$ is trivially defined as
$f(\mathcal{A})=\sum_j {\cal A}_{ij}={\bf k}$ (where by construction ${\bf k} \in \Omega$), the challenging part and thus the strategy of this proof is to construct $f^{-1}$. In order to do that we first propose an algorithm that maps the degree sequence $\textbf{k}$ into a certain matrix $\mathcal{B}$ and accordingly we prove that $\mathcal{B}=\mathcal{A}$.
To further prove this latter part, we show that $\mathcal{B}$ is an adjacency matrix of the same order as $\mathcal{A}$, then we show that every edge in $\mathcal{B}$ is also in $\mathcal{A}$, and finally we show that the number of edges in $\mathcal{B}$ is the same as the number of edges in $\mathcal{A}$.\\

\noindent Consider a HVG of order $N$ with degree sequence $\bf k$, and consider a replica of this degree sequence which we will update, where initially ${\bf k}^{(0)}:=[k_1,k_2,\dots,k_N]$. Then $f^{-1}({\bf k})\equiv {\cal B}$ is an element from the set $\{0,1\}^{N\times N}$ which is constructed as follows:\\

\begin{figure}
\centering
\includegraphics[width=1.0\columnwidth]{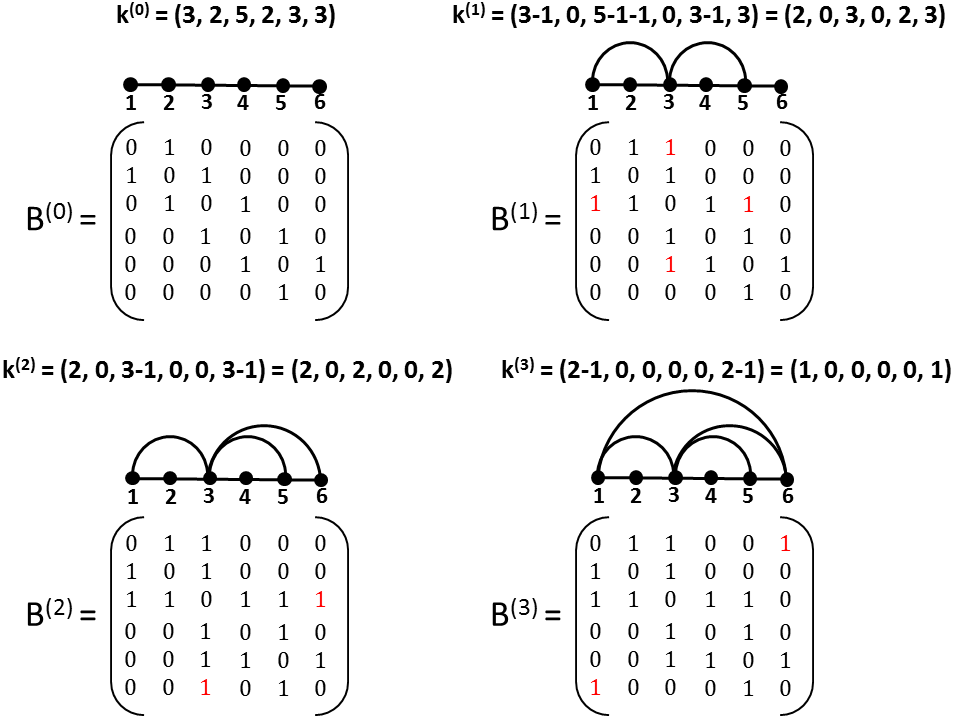}
\caption{Example of how we can reconstruct the HVG (i.e. the adjacency matrix) from the degree sequence ${\bf k}=(k_1,\dots,k_N)$ from the canonical series of figure \ref{fig:canoni}, where $N=6$ and ${\bf k}=(3,2,5,2,3,3)$. In the setting, we always have a Hamiltonian path. At each step of the algorithm, we locate the nodes with degree two and we decimate them (see the text), associating a link between the neighbors of the nodes with degree 2. The process runs iteratively until we find the final sequence $(1,0,\dots,0,1)$.}
\label{fig:cartoon}
\end{figure}

\noindent \texttt{Setting:}\\
Start setting ${\cal B}_{ij}=0 \  \forall i,j$, and ${\cal B}_{i,i+1}={\cal B}_{i+1,i}=1 \ \forall i=1,\dots,N-1$.  These links are obviously also present in $\mathcal{A}$ since $\bf k$ is associated to a HVG and then this graph has the trivial Hamiltonian path $1-2-3\dots-N$. An illustration of how the algorithm works in a concrete example is shown for illustration in figure \ref{fig:cartoon}.\\

\noindent \texttt{Step 1 (Initial assignment of $k=2$ nodes):}\\
Locate in ${\bf k}^{(0)}$ those inner vertices $j$ (where $j \in [2,N-1]$) with degree $k_j=2$. Note that this is always possible: the original time series $\{x_i\}$ has a finite number of elements, hence we can always find its minimum datum. This minimum is not in the boundary ($j\neq 1,N$) as the series is canonical (the only pathological case where this cannot be done is when the time series is constant, which according to proposition \ref{prop} means that the HVG is the path graph). By construction, this minimum will be associated to a vertex in the HVG with degree $2$.\\
For each of these inner vertices, the associated data will be a local minimum (meaning that at least $x_j<x_{j-1},x_{j+1}$), therefore there exists an edge between the neighbors of $j$ in  ${\cal A}$. We update the adjacency matrix ${\cal B}_{j-1,j+1}={\cal B}_{j+1,j-1}=1 \ \forall j$ accordingly. To carry track of the number of edges that we are introducing in ${\cal B}$, we also update ${\bf k}^{(0)}\to{\bf k}^{(1)}$ by removing there the edges that were introduced in ${\cal B}$. Thus $\forall j$ such that $k_j=2$, $k_j=2\to 0$, $k_{j+1}\to k_{j+1}-1$ and $k_{j-1}\to k_{j-1}-1$.\\
After step 1 the series has been decimated, and some edges have accordingly been deleted in ${\bf k}^{(\cdot)}$. Incidentally, note that this process had been used previously in the context of a graph-theoretical renormalization group transformation \cite{jns}.\\

\noindent \texttt{Step $>1$ (Iteration):}\\
We further locate in ${\bf k}^{(1)}$ those vertices $j$ that have been updated and now have $k_j=2$. For each of these vertices, we locate the closest vertices $r$ and $l$ (where $r>j$ and $l<j$) for which $k_r>0$ and $k_l>0$. These are the new neighbors in the decimated series and accordingly ${\cal B}_{rl}={\cal B}_{lr}=1$. Note that by construction, we have $x_p<\inf(x_r,x_l) \ \forall p:l<p<r$. This holds because all the data placed between $l$ and $r$ are either $j$ or data that were already decimated. Therefore this new edge was also present in the original adjacency matrix, ${\cal A}_{rl}={\cal A}_{lr}=1$. Again, we carry track of these new edges by updating ${\bf k}^{(1)}\to{\bf k}^{(2)}$ such that $k_j=2\to 0$, $k_{r}\to k_{r}-1$, and $k_{l}\to k_l-1$.\\
We repeat this last step iteratively (this is always possible as initially all data are different so the process of finding the smaller data is always possible). This iteration will set at each step the $k_j=2$ vertices to zero in the updating degree sequence and will decrease by one the value of their neighbors.  The process stops after we have repeated this process an unknown number of times $m$ and reach ${\bf k}^{(m)}=[1,0,\dots,0,1]$ which is the trivial absorbing state. Once this limit is reached the algorithm stops and we introduce the last edge, setting ${\cal B}_{1N}={\cal B}_{N1}=1$ (which is also a link present in ${\cal A}$ as we are dealing with a canonical HVG) and update the degree sequence accordingly ${\bf k}^{(m+1)}=[0,\dots,0]$. No more edges can be deleted in the degree sequence and no more edges can be added to ${\cal B}$.\\

\noindent All in all, this procedure uses ${\bf k}^{(\cdot)}$ to construct a $[0,1]^{N\times N}$ matrix ${\cal B}$. As we have shown, all the new edges introduced in ${\cal B}$ (all the 1s) also belong to ${\cal A}$. On the other hand, the total number of elements removed from ${\bf k}^{(0)}$ is exactly twice the total number of edges that were introduced in ${\cal B}$ (if we take into account all steps in the algorithm above). Thus the number of edges introduced in ${\cal B}$ is $\frac{1}{2}\sum_{i=1}^N k_i=E$, where $E$ is the number of edges in the original HVG. Therefore there are no missing edges and ${\cal B}={\cal A}$. \hfill $\blacksquare$\\

\noindent {\bf Concluding remarks. }
To be able to put in bijection a graph's adjacency matrix with its degree sequence is a nontrivial result, as the adjacency matrix stores in principle $N^2$ entries, while the degree sequence only requires $N$ of them. This optimal compression property is a very particular property of unigraphs, in this sense a straightforward corollary of our main result states that canonical HVGs are unigraphs, and thus this family of HVGs shares all their nice properties such as having linear recognition time (something that was proved independently for HVGs recently \cite{severini}). Note that the converse is not true, as it can be trivially proved by counterexample or by the fact that HVGs are outerplanar and not all unigraphs are outerplanar. It should be highlighted that the canonicity condition can in some cases be relaxed, in the sense that in many instances the bijection described above works even for non-canonical HVGs. However for the theorem to hold we needed to restrict the set of HVGs to those that are canonical. We suspect that for most of the practical situations the theorem still holds, however this remains an open problem.\\
Nevertheless the theorem implies that all the information encoded in the graph is efficiently compressed in its degree sequence, giving a satisfying solution to the paradox of the unreasonable efficiency of this simple topological metric. This finding should allow HVG-based feature extraction algorithms to safely focus on the degree sequence, saving both memory allocation and execution time without losing information.


\bibliography{apssamp}

\end{document}